\documentclass[aps,prl,twocolumn,floats,prl,nofootinbib,superscriptaddress]{revtex4-1}

\usepackage{graphicx,amsmath,amsfonts,amssymb,slashed}
\usepackage[usenames,dvipsnames]{xcolor} 

\usepackage{graphicx}
\usepackage{dcolumn}
\usepackage{xcolor}
\usepackage{bm}
\usepackage{natbib}
\usepackage{calligra}
\usepackage[T1]{fontenc}
\usepackage{egothic}
\usepackage[T1]{fontenc}
\newfont{\rsfsten}{rsfs10 scaled 1200}
\newfont{\rsfsseven}{rsfs10 scaled 1200}
\newfont{\rsfsfive}{rsfs10 scaled 1200}
\usepackage{epsfig}
\usepackage{units}
\usepackage[utf8]{inputenc}
\usepackage{multirow}

\begin{document}

\title{Probing Primordial-Black-Hole Dark Matter with Gravitational Waves}

\author{Ely D. Kovetz} 
\affiliation{
Department of Physics and Astronomy, Johns Hopkins
     University, 3400 N.\ Charles St., Baltimore, MD 21218, USA}

\begin{abstract}
Primordial black holes (PBHs) have long been suggested as a candidate for making up some 
or all of the dark matter in the Universe. Most of the theoretically possible mass range for PBH 
dark matter has been ruled out with various null observations of expected signatures of their 
interaction with standard astrophysical objects. However, current constraints are significantly 
less robust in the $20\,M_\odot \lesssim M_{\rm PBH} \lesssim 100\, M_\odot$ mass window, 
which has received much attention recently, following the detection of merging black holes 
with estimated masses of $\sim30\,M_\odot$ by LIGO and the suggestion 
that these could be black holes formed in the early Universe. We consider the potential of 
advanced LIGO (aLIGO) operating at design sensitivity to probe this mass range 
by looking for peaks in the mass spectrum of detected events. To quantify the background, 
which is due to black holes that are formed from dying stars, we model the shape of the 
stellar-black-hole mass function and calibrate its amplitude to match the O1 results. Adopting very 
conservative assumptions about the PBH and stellar-black-hole merger rates, we show that $\sim5$ years of aLIGO data can 
be used to detect a contribution of $>20\,M_\odot$ PBHs to dark matter down to $f_{\rm PBH}<0.5$ 
at $>99.9\%$ confidence level. Combined with other probes that already suggest tension 
with $f_{\rm PBH}\!=\!1$, the obtainable independent limits from aLIGO will thus enable a firm test 
of the scenario that PBHs make up all of dark matter. 
\end{abstract}

\maketitle

One of the cornerstones of $\Lambda {\rm CDM}$, the concordance cosmological 
standard model, is the cold dark matter (DM) component that makes up $\sim\!25\%$ 
of the energy density in the Universe today. While the evidence for its existence are compelling 
\cite{Ade:2015xua,Freese:2017idy}, the nature of it is still unknown. 
As the limits on models of particle dark matter (in particular weakly-interacting 
massive particles, known as WIMPs \cite{Jungman:1995df}) are tightening 
\cite{Essig:2012yx,Aprile:2012nq,Akerib:2015rjg}, it is becoming ever more important to 
consider alternative models.

An especially intriguing candidate to make up the invisible form of matter in the Universe is 
primordial black holes (PBHs), which are black holes that are formed deep in the 
radiation era of the infant Universe \cite{Carr:1974nx,Meszaros:1974tb,Chapline:1975,Carr:1975qj,Clesse:2015wea}. 
Based on various observations, the contribution of PBHs to dark matter has been strongly 
constrained across more than 30 orders of magnitude of their theoretically possible mass 
range \cite{Carr:2016drx,Kuhnel:2017pwq,Carr:2017jsz}. Still, in several mass windows 
existing constraints are less stringent and additional probes are called for.
This is especially true for the $20\,M_\odot \lesssim M_{\rm PBH} \lesssim 100\, M_\odot$ 
 window, which has attracted much interest as a result of the first detection of 
merging black holes with measured masses of $\sim30\,M_\odot$ by the LIGO observatory 
\cite{Abbott:2016blz}, following the demonstration in Ref.~\cite{Bird:2016dcv} that the 
predicted merger rate for PBHs in this mass range is consistent with the estimated 
event rate for high mass mergers from the O1 aLIGO data \cite{TheLIGOScientific:2016pea}.

\begin{figure}
\includegraphics[width=\columnwidth]{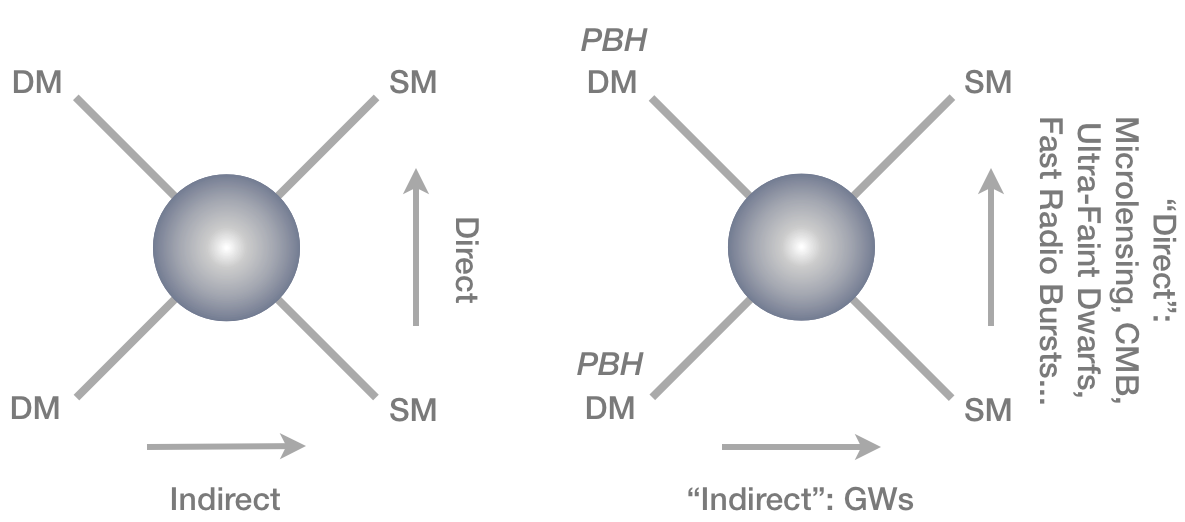}
\vspace{-0.25in}
\caption{Hunting for dark matter: the left panel is the popular illustration of the different search 
methods for dark matter particles. One avenue is {\it direct} detection, whereby dark matter particles are targeted 
by looking for their effect on standard model particles (e.g.\ recoil on heavy nuclei in underground 
experiments). The second is known as {\it indirect} detection, where the goal is to observe the 
products of dark matter self-interaction (e.g.\ gamma-rays produced from annihilating WIMPs \cite{Gaskins:2016cha}). 
Analogously, we group all the methods focusing on the effect of PBHs on standard astrophysical 
objects as ``direct'' detection (including probes such as microlensing \cite{Alcock:1996yv,Tisserand:2006zx,Mediavilla:2017bok}, CMB anisotropies \cite{Ricotti:2007au,Ali-Haimoud:2016mbv,Blum:2016cjs}, dynamical heating of ultra-faint dwarf galaxies \cite{Brandt:2016aco,Koushiappas:2017chw}, number counts of compact X-ray objects \cite{Inoue:2017csr}, etc.;\ and in the future, strong lensing of FRBs \cite{Munoz:2016tmg} and pulsar-timing \cite{Schutz:2016khr}). 
``Indirect'' detection of PBH dark matter involves looking for gravitational waves 
emitted when a PBH pair is ``annihilated'' by merging into a larger BH \cite{Abbott:2005pf}.} 
\label{DMsearchmethods}
\vspace{-0.25in}
\end{figure}

One could describe the search for PBH dark matter in analogy to the one for particle dark matter, 
as is illustrated in Fig.~\ref{DMsearchmethods}. 
The constraints on the former to date have been solely based on ``direct'' detection
searches, involving possible interactions between PBHs and standard astrophysical objects. 
In this {\it Letter}, we  consider the prospects of the ``indirect'' search path for PBH dark matter, 
namely the production of standard (cosmological) model signals in the form of 
gravitational waves as a result of PBH self-interaction (or ``annihilation''). 
The key to this approach is to understand and quantify the background as well as possible, and to identify unique 
features in the dark matter signal that can tell them apart.
 
Examples of such features are the orbital eccentricity of the coalescing binary and the black hole spins. 
The former was investigated recently in Ref.~\cite{Cholis:2016kqi}, but unfortunately
the prospects for detecting events with a non-zero trace of the initial eccentricity in aLIGO 
are quite dim. As for the spin, the problem is twofold. The initial spin distribution of PBHs is unclear 
on the one hand (see \cite{Chiba:2017rvs} for some estimates), and on the other, 
the key identifier of PBH mergers in this regard---that the spin of the two black holes should not be aligned---is 
mimicked by various models of dynamical binary formation \cite{Rodriguez:2016vmx,Talbot:2017yur} 
(whose rate is currently very uncertain), and so it is very hard to distinguish them from the background.

Here we focus on the mass spectrum of merging black hole binaries as a 
probe of PBH dark matter. The idea is simple: if PBHs in a given mass range account
for all (or some fraction) of the dark matter in the Universe, we should see an excess of merger
events involving black holes in the corresponding mass bin \cite{Kovetz:2016kpi}. Provided that this 
excess is large enough to be differentiated from the expected background from mergers of black holes 
of stellar origin, a detection, or lack thereof, could either support the existence or place limits on the 
abundance of PBHs. This abundance is parameterized by the quantity $f_{\rm PBH}$, the fraction 
of dark matter in PBHs.

Our goal is to provide the 
most reliable constraints on PBH dark matter---focusing especially on the most motivated question which is whether all of 
dark matter can be explained by PBHs---and we therefore choose to be very careful and consider the 
most pessimistic case for PBH dark-matter detection, i.e.\ the {\it lowest} estimated rate of PBH mergers 
and the {\it highest} rate of stellar black hole mergers. We will show that even in this challenging case, 
gravitational-wave observations by aLIGO at design sensitivity should within a decade either exhibit strong hints for 
a PBH contribution to dark matter, or rule them out as the single form of dark matter (albeit allowing 
for a considerable fraction of it to made up of PBHs). 

To begin, we review the suggested mechanisms of PBH binary formation. 
The first model was put forward in Refs.~\cite{Nakamura:1997sm,Sasaki:2016jop}, where it was demonstrated 
that a subset of an initial PBH population, assumed to be randomly distributed in space, 
would be in close enough proximity to overcome the cosmic expansion and form bound pairs. 
The distribution of the semi-major axis of these binaries will be quite wide. However, 
as they generically have high initial orbital eccentricity (they avoid a head-on collision due to the 
influence of the closest neighbors), some will have merger times that allow them to reach the 
endpoint of their coalescence within the detectable volume of aLIGO. 
Ref.~\cite{Sasaki:2016jop} predicts a very high merger rate (which is in fact already in 
tension with existing observations if $f_{\rm PBH}\gg0.01$).  
However, a crucial question is whether these early-formed binaries can survive as bound pairs 
throughout the evolution of the Universe, without being disrupted. This was briefly addressed  
in Ref.~\cite{Sasaki:2016jop}, where the probability for disruption was calculated to be as small as $\mathcal{O}(10^{-7})$, 
but only for a binary that resides in a Milky-Way type halo today. Since PBHs binaries in this scenario were created 
very early on, before the formation of large dark matter halos (in a series of violent cosmic processes), this 
is at best an underestimate of the actual disruption rate. A thorough reexamination of this model, taking 
into account the interaction of PBH binaries with other PBHs, the rest of dark matter (if $f_{\rm pbh}\ll1$) and with baryonic matter, 
is under way, but is outside the scope of this {\it Letter}. In short, Ref.~\cite{Ali-Haimoud:2017} evaluates the chance for disruption 
by tidal effects of the smooth halo and encounters with other PBHs (which are more efficient in the first halos, which are denser 
and have lower velocity dispersions), finding that it is  higher than previously estimated. Meanwhile, the effect of a circumbinary 
accretion disk has been examined in Ref.~\cite{Hayasaki:2009ug}, concluding that it could lead to a decrease in the semi-major 
axis on a timescale fast enough such that all early-formed binaries with masses in the stellar-mass range will have merged well 
before redshift $z\sim1$, and therefore remain outside the reach of aLIGO. Aiming to provide the most robust bounds on PBH dark 
matter, we shall therefore treat this rate as an optimistic case, and proceed to focus on more conservative scenarios, described below.

A second model of PBH binary formation was proposed in Ref.~\cite{Bird:2016dcv}. In this model, PBHs form binaries in 
close two-body encounters, as a result of energy loss from gravitational-wave emission as they pass each 
other by. The rate for this process to occur was calculated for dark matter halos of different masses (and densities 
and velocity dispersions), and when integrated over a full mass function (cutting off at the low mass end where 
halos would be too small not to evaporate by dynamical relaxation), the total merger rate was found to be
\begin{equation}
R_{\rm PBH}\approx2f_{\rm PBH}^{53/21}\left(M_{\rm PBH}/30M_\odot\right)^{-11/21}\,{\rm Gpc^{-3} yr^{-1}},
\label{Rpbh}
\end{equation}
where we have used the fact that the evaporation time is primarily determined by the number of black holes in the 
halo, as their density in this limit---and thus the dynamical time---is roughly constant with mass.
This rate for PBHs with mass $M_{\rm PBH}=30\,M_\odot$ is consistent with the recent LIGO $90\%$-interval 
estimate for black holes as massive as the ones in the first detection, $0.5-12\,{\rm Gpc^{-3} yr^{-1}}$ \cite{TheLIGOScientific:2016pea}.
Of the many assumptions in this calculation, by far the most daring is the extrapolation of the halo mass function, 
velocity dispersion distribution, density profile and mass-concentration relation to very low ($\sim10^3\,M_\odot$) dark matter 
halo masses, orders of magnitude below what can be observed or even simulated. The good news however, is that as 
these binaries have a very small separation when they form, they merge very quickly and are not susceptible to the effect of
interfering processes such as mentioned above for early-formed binaries. While a bias against a 
more-optimistic higher rate is not too worrisome for the purposes of this work, it is more crucial to have a clear idea for the 
lowest reasonable bound on the merger rate. In work to appear \cite{Nishikawa}, a merger rate is calculated for the same
two-body formation mechanism of PBH binaries, but focusing on the contribution to the merger rate from encounters
that occur inside dark matter density spikes around supermassive black holes (SMBHs) at the centers of galaxies 
\cite{Gondolo:1999ef}. Integrating over the mass function of SMBHs from $\sim10^5\,M_\odot$ to $\sim10^9\,M_\odot$, 
this is found to yield roughly $>10\%$ of the total rate from all halos, providing a highly-conservative lower floor for the total PBH merger rate.  

We will therefore proceed by adopting $R_{\rm PBH}$ in Eq.~(\ref{Rpbh}) as a conservative estimate for the 
rate of PBH mergers in the local Universe, addressing the pessimistic case of a rate ten times lower, and bearing in 
mind that if early-Universe binaries are somehow found to be stable to disruption, the rate could be order(s) of magnitude higher.

The next step is to make a prediction for the background, which consists of 
mergers of stellar black holes that contribute to the mass spectrum of detected events. To do this, we require an 
assumption for the astrophysical merger rate (which depends on factors such as metallicity and merger time-delay
distributions) and for the mass function of stellar black holes.
For the merger rate, we follow Ref.~\cite{TheLIGOScientific:2016htt}. We use a simple approximation, 
$R(z)\simeq97(1+z)^2\,{\rm Gpc^{-3} yr^{-1}}$, which provides a good fit to their results at redshifts $z<1$ \cite{TheLIGOScientific:2016pea}. 
As for the black hole mass function, we follow Ref.~\cite{Kovetz:2016kpi} and consider a simple ansatz whereby the 
probability distribution function (PDF) of the black hole mass is described by a simple power law (motivated by the slope 
of the stellar initial mass function, which has been corroborated by numerous observations in the $1-100\,M_\odot$ mass
range \cite{Salpeter:1955it,Kroupa:2000iv}). We impose sharp and exponential cutoffs at the lower and upper end 
of the mass spectrum, respectively, to take into account the neutron-star---BH transition threshold and the increasing 
wind-driven mass loss of high mass stars in the Wolf-Rayet phase. Denoting the black hole mass by $M_1$, the PDF is 
given by
\vspace{-0.01in}
\begin{equation}
P(M_1)= A_{M_1}M_1^{-\alpha}\mathcal{H}(M_1\!-\!M_{\rm gap})e^{-M_1/M_{\rm cap}},
\label{eq:BHMF}
\end{equation}
where $\mathcal{H}$ is the heaviside function and $A_{M_1}$ is an overall normalization. 
We shall use as fiducial values $\alpha=2.35$ (following \cite{Kroupa:2000iv}), $M_{\rm gap}=5\,M_\odot$ 
(motivated by current observations \cite{Bailyn:1997xt,Ozel:2010su,Farr:2010tu} and by some theoretical works 
\cite{Belczynski:2011bn,Fryer:2011cx,Kochanek:2013yca}) and $M_{\rm cap}=60\,M_\odot$. 
The latter choice naturally affects our constraints strongly at masses beyond the cutoff, as with a vanishing background, 
PBHs of that mass would be easier to detect. However, very massive PBH binaries will have a merger frequency too 
small to be detected by aLIGO, which offsets this effect. We deliberately avoid choosing a lower value, to reflect our 
ignorance and so as not to artificially strengthen the bound on $f_{\rm PBH}$. With future observations, 
it might be possible to improve on these somewhat arbitrary choices, perhaps leading to more tightened constraints. 

Our observable is the total number of detected events as a function of the black hole mass. In order to 
determine the background contribution, we need to model the mass ratio between the black hole binaries 
as well. Setting $M_1$ from now on to be the mass of the heavier black hole in each binary, we follow 
Ref.~\cite{TheLIGOScientific:2016pea} and assume that the mass of the lighter black hole, $M_2$, 
has a uniform distribution in the range $[M_{\rm gap},M_1]$, given by
\begin{equation}
P(M_2)= A_{M_2}\mathcal{H}(M_2\!-\!M_{\rm gap})\mathcal{H}(M_1\!-\!M_2).
\label{eq:pm2}
\end{equation}

The observable redshift volume of gravitational waves from BH mergers depends on the instrumental properties.
The signal-to-noise ratio ($S/N$) for a single interferometer detector is given by
$({S/N})^{2} =  \int_{f_{\rm min}}^{f_{\rm max}}
df {4h_{c}^{2}(f)}/5{S_{n}(f) (2 f)^{2}}$,
where $h_{c}(f)$ is the observed strain amplitude and $S_{n}(f) = h_{n}^{2}(f)$ 
is the strain noise amplitude (for more details, see Ref.~\cite{Cholis:2016kqi} and references therein). 
We follow convention and set the detection threshold at $S/N>8.0$ \cite{Abadie:2010cf}. 
We use the approximated analytical model for aLIGO noise of Ref.~\cite{Ajith:2011ec}, setting 
$f_{\rm min}=10\, {\rm Hz}$, above which the curve matches the official LIGO
curve very well \cite{Shoemaker2010}. 

Lastly, inferring the mass of the merging black holes from the gravitational-wave signal involves an associated uncertainty. 
We model this by convolving the mass function with a log-normal distribution reflecting a $5\%$ relative mass error for 
aLIGO observations (see Ref.~\cite{Kovetz:2016kpi}). This choice suggests a minimal width for our binning of the mass function 
when calculating the signal-to-noise.

We are now equipped to make a theoretical prediction for the total number of detected background 
merger events over a time $T_{\rm obs}$ with a given mass $M_1$ (the mass of the heavier BH in the one-dimensional (1D) 
case), or two masses $M_1,M_2$ (in the 2D case), which is given by
\begin{eqnarray}
\frac{dN^{\rm BG}(M_1)}{dM_1} &=& 4\pi P({M_1})T_{\rm obs} \int\limits_{M_{\rm gap}}^{M_1} P(M_2)dM_2  \cr 
&&\times\int\limits_0^{z_{\rm max}(M_1, M_2)}\frac{c\chi(z)^2R(z)}{(1+z)H(z)}dz.
\label{eq:dN}
\end{eqnarray}
Here the observable redshift volume is defined by $z_{\rm max}$, the 
maximum redshift up to which a BH merger with masses $M_1,M_2$ can be detected; 
$H(z)$ is the Hubble parameter and $\chi(z)$ is the radial comoving distance. 
In the 2D case, we use $dN(M_1,M_2)/dM_1dM_2$, dropping the first integration in  Eq.~(\ref{eq:dN}). 
Integrating within each mass bin $i$ with edges $[M_{\textrm{min},i},M_{\textrm{max},i}]$, we finally get
\begin{equation}
N^{\rm BG}_i = \int_{M_{\textrm{min},i}}^{M_{\textrm{max},i}}\frac{dN(M_1)}{dM_1}dM_1.
\label{eq:N}
\end{equation}
We then divide $N(M_1)$ into $50$ logarithmic bins from $4\,M_\odot$ to $120\,M_\odot$
and likewise $N(M_1,M_2)$ into 100 bins (10 along each  axis). 
The bin width is chosen such that the mass measurement error is subdominant, and the variance in 
each bin is set by the Poisson error $\sigma_i^2=N_i$. 

The most powerful way to constrain the abundance of a PBH population with a narrow mass 
distribution is to look at the 2D mass distribution of detected events, and check for a peak in the 
mass bin(s) surrounding the central value of that distribution. This approach makes use of all 
the available data, but since the background needs to be calculated for each mass bin, making
accurate forecasts requires a good understanding of the black-hole mass function as well as 
the binary mass ratio. To somewhat relax this model dependence, we can also choose to limit 
ourselves to using only the number counts for the heavier BH in each binary (the mass-ratio 
will still enter the signal-to-noise calculation through the fiducial choice of the $M_2$ PDF, 
but since each column in the one-dimensional $M_1$ distribution is an integral quantity over 
the full $M_2$ mass range, the dependence on this choice is weaker in the 1D case).
The widely used convention when forecasting limits on the fraction of dark matter in PBHs is to 
assume a delta-function PBH mass function. In practice, we  assume Gaussian PDFs for 
$M_{\rm PBH}=M^{\rm PBH}_1=M^{\rm PBH}_2$ with a $5\%$ width (the precise choice is 
unimportant as long as this width is smaller than the measurement error). Together 
with the rate in Eq.~(\ref{Rpbh}), we get $N^{\rm PBH}_i(f_{\rm PBH}, M_{\rm PBH})$ for each value 
of $f_{\rm PBH}$ and $M_{\rm PBH}$ using the prescription in Eqs.~(\ref{eq:dN}),(\ref{eq:N}).

\begin{figure}
\includegraphics[width=\columnwidth]{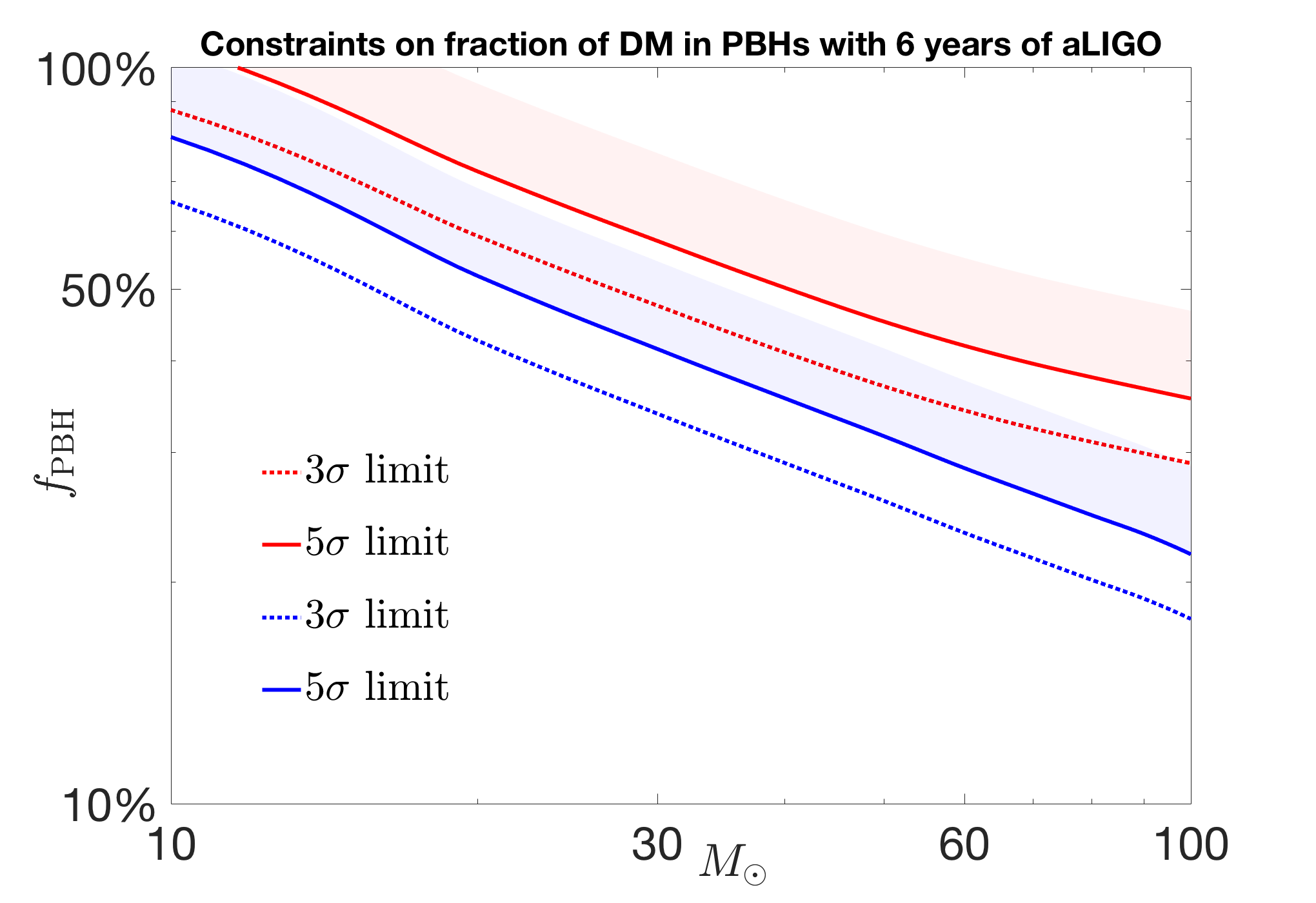}
\vspace{-0.3in}
\caption{Constraints on the fraction of dark matter in primordial black holes, as a function of their mass. 
It is conventional to assume a delta function PBH mass function when calculating constraints such as these. 
In practice, however, the measurement error will widen the observed distribution, so we model the PBH 
mass function as a Gaussian with a $5\%$ width. The dashed lines correspond to 3$\sigma$ limits 
(i.e.\ requiring that the fraction of DM in PBHs is low enough such that the number of PBHs not exceed 
the Poisson noise of the stellar mergers by more than a factor of 3) and the solid lines are more stringent 
$5\sigma$ limits. Constraints based on the 1D mass distribution are shown in red, those using the 2D 
distribution (which are tighter as the noise per bin is smaller) in blue. For $30\,M_\odot$, we get 
$f_{\rm PBH}\lesssim50\%$ at $5\sigma$ ($3\sigma$) in the 2D (1D) case. The bands above the solid 
lines extend up to a factor $400\%$ ($200\%$) uncertainty in the background (signal) amplitude.}
\label{PBHConst}
\vspace{-0.2in}
\end{figure}

We now derive our forecast for the limits that aLIGO at design sensitivity can impose with the planned 
$6$ years of observation by solving the following equation for $f_{\rm PBH}$
\vspace{-0.13in}
\begin{equation*}
S/N=\sqrt{\sum\limits_i\left(\frac{N^{\rm Tot}_i-N^{\rm BG}_i)}{\sqrt{N^{\rm Tot}_i}}\right)^2};~N^{Tot}\equiv N^{\rm PBH}_i+N^{\rm BG}_i
\end{equation*}
\vspace{-0.13in}
\begin{equation}
\longrightarrow\sqrt{\sum\limits_i\left(\frac{N^{\rm PBH}_i(f_{\rm PBH}, M_{\rm PBH})}{\sqrt{N^{BG}_i}}\right)^2}-n_\sigma=0,
\label{bound}
\end{equation}
where we set a desired signal-to-noise ratio of $n_\sigma=3$ or $5$ standard deviations and have assumed under the 
null hypothesis ($N_{\rm PBH}=0$) that the (Poisson) error in each bin is  $\sqrt{N^{\rm BG}_i}$ (using the Gaussian approximation 
for the Poisson distribution is valid given the number of events in each bin).
The result---in the form of $3-$ and $5\!-\!\sigma$ limits on $f_{\rm PBH}$ for each PBH mass---is shown in 
Fig.~\ref{PBHConst}. We see that based on the rate in Eq.~(\ref{Rpbh}), 
the scenario in which all the dark matter is in the form of PBHs can be strongly tested (ruled out at $\gg5\sigma$) 
when using either the full 2D or 1D mass spectra of observed BH mergers. 

Note that we have assumed the stellar-BH mass-function parameters can be held fixed, rather than 
fitting for them in tandem with the amplitude of the PBH contribution and marginalizing over them. Under 
the assumption that it is smooth, consistent with \cite{TheLIGOScientific:2016pea}, the effect of this 
approximation on our results is small. We emphasize that if it is found to deviate strongly from Eq.~(\ref{eq:BHMF}), future data from planned next generation GW experiments such as Cosmic Explorer \cite{Evans:2016mbw}, Einstein Telescope \cite{ET}, DECIGO \cite{Seto:2001qf} and LISA \cite{2017arXiv170200786A}---which will be sensitive enough at low frequencies to be able to detect BH mergers at high redshifts, well beyond the peak of the star-formation rate--- will allow a straightforward discrimination between the stellar and primordial merging-BH populations, based on their very different redshift distributions \cite{Koushiappas:2017kqm}. 
Another source of uncertainty is the overall background amplitude, whose current $90\%$-confidence range 
still spans roughly an order of magnitude \cite{TheLIGOScientific:2016pea}.

To incorporate the various modeling uncertainties, we use the dependence of Eq.~(\ref{Rpbh}) on $f_{\rm pbh}$ and of Eq.~(\ref{bound}) on 
the background and PBH rates, to show in Fig.~\ref{PBHConst} a band encompassing an underestimate (overestimate) of up to a factor of $400\%$ ($200\%$) in the 
background (signal) rates. As can be seen, our conclusion that the scenario of PBH DM can be tested convincingly ($\gg5\sigma$)  
in the range $20\,M_\odot \lesssim M_{\rm PBH} \lesssim 100\, M_\odot$ is quite robust.
even with a PBH merger rate ten times lower, which as described above should be
taken as an ultra-conservative bound, $f_{\rm PBH}=1$ can still be rejected at $5\sigma$ confidence across the $20-100\,M_\odot$ range 
using the 2D information.
Naturally, if the rate is much higher \cite{Sasaki:2016jop,Ali-Haimoud:2017}, a null detection will yield even stronger limits on $f_{\rm PBH}$, motivating efforts to better understand the mechanisms of PBH binary formation and disruption, using both analytic methods and simulations. 
Eventual bounds will depend on additional real-world factors, such as the operating sensitivity of aLIGO during its six year run. 
In particular, we stress that the constraints can be {\it significantly improved} if the signal-to-noise threshold for a detected event in 
each interferometer is reduced, as should be done when performing a statistical analysis of an ensemble of events \cite{Kovetz:2016kpi}.

While the constraints we forecast are weaker than some that have been already claimed 
or projected by other methods, they constitute a unique and independent method of testing the scenario, by focusing on the self-interaction 
of PBHs rather than their interaction with other astrophysical objects. They are subject to different systematics and 
modeling assumptions, and since the analysis presented here {\it heavily} errs on the side of caution and still finds 
promising results, they represent a truly robust test, achievable within a decade, of the important cosmological scenario
that dark matter is made of PBHs. 

We thank Yacine Ali-Ha\"imoud, Simeon Bird, Marc Kamionkowski and especially Ilias Cholis for discussions. This work was supported by NSF Grant
No. 0244990, NASA NNX15AB18G and the Simons Foundation.

\end{document}